\documentclass[aps,prc,showpacs,12point]{revtex4}

\usepackage{graphicx,amsmath,amssymb,bm}

\newcommand{\eq}[1]{Eq.~(\ref{#1})}
\newcommand{\be}{\begin{equation}}
\newcommand{\ee}{\end{equation}}
\newcommand{\bea}{\begin{eqnarray}}
\newcommand{\eea}{\end{eqnarray}}

\newcommand{\psidagop}{\hat{\psi}^\dagger}
\newcommand{\psiop}{\hat{\psi}}

\def\bfq {{\bf q}}

\def\bfK{{\bf K}}

\def\bfp{{\bf p}}  
\def\bfx{{\bf x}}\def\bfy{{\bf y}}\def\calU{{\cal U}}
\def\bfz{{\bf z}}
\begin{document} 

\preprint{NT@UW-08-16}
 
\title{NT@UW-08-16\\Understanding the Optical Potential in HBT Interferometry}

\author{Matthew Luzum, John G.~Cramer and Gerald A.~Miller}

\affiliation{Department of Physics, University of Washington,
Seattle, WA 98195}
\email[E-mails:~]{mluzum@phys.washington.edu, cramer@phys.washington.edu,
miller@phys.washington.edu}

\date{\today}

\begin{abstract} 
The validity of using a pion optical potential to incorporate the
effects of final state interactions on HBT interferometry is investigated.
We find that if 
the optical potential is real, the standard formalism is modified as previously described in the literature.
 However, if the optical potential is complex, 
a new term involving pion emission from  
eliminated states must be included. The size
of such effects in previous work by Cramer and Miller is assessed.
\end{abstract} 


\maketitle

\section{Introduction}

\large
The space-time structure of the 
``fireball'' produced
in the collision between two relativistically  moving heavy ions can be investigated
by  measuring  two-particle momentum correlations
between pairs of identical bosons.  The  Bose-Einstein enhancement of the 
coincidence rate at small momentum differences
depends on  the space-time extent  of the particle source.
This method of investigation,
 called HBT interferometry,  has been applied extensively 
in recent experiments at the Relativistic
 Heavy Ion Collider (RHIC) by the
STAR and PHENIX 
collaborations. See the reviews\cite{Pratt:wm,Wiedemann:1999qn,Kolb:2003dz,Lisa:2005dd}.

The invariant ratio of the 
 cross section for the production of two 
  pions of momenta $\bfp_1,\bfp_2$
to the product of single particle production cross sections is analyzed as
the correlation function $C(\bfp_1,\bfp_2)$.
 We define
 $\bfq$=$\bfp_1$--$\bfp_2$
and $\bfK$=$(\bfp_1$+$\bfp_2)/2$, with $\bfK_T$ as 
the component perpendicular to the beam direction. 
(We focus on mid-rapidity data, where 
$\bfK=\bfK_T$.)
The correlation function
can be   parameterized
for small $\bfq$ as 
$
C(\bfq,\bfK)-1 \approx \lambda \exp{(-R_O^2q_O^2-R_S^2q_S^2-R_L^2q_L^2)}\approx
\lambda(1-R_O^2q_O^2-R_S^2q_S^2-R_L^2q_L^2)\; (q_iR_i\ll1), $   
where $O,S,L$ 
represent directions parallel to $\bfK_T$, perpendicular
to both $\bfK_T$ and the beam direction, 
and parallel to the beam direction \cite{BP-HBT}. 
Early \cite{Rischke:1996em} and recent \cite{Kolb:2003dz} hydrodynamic 
calculations predicted  
that  a fireball evolving through a quark-gluon-hadronic phase 
transition would emit pions
over a long time period, causing a large ratio $R_O/R_S$.
The puzzling experimental
result that $R_O/R_S\approx 1$ 
\cite{Adler:2001zd} is
part of what has been  
 called ``the RHIC HBT puzzle'' \cite{Heinz:2002un}. 
 Another part of the puzzle is that the measured radii depend strongly
on the average momentum $K$, typically decreasing in size by about 50\% over
the measured range,  
 showing that the radii are not simply a property of a static 
source.
The medium at RHIC seems 
to be a very high density, strongly interacting plasma \cite{Gyulassy:2004zy},
so that any pions made in its interior could be expected to interact strongly before emerging.
Thus one expects that the  influence 
of the interactions between the pion probe and the medium, 
as well as flow and other effects,
must be taken into account when extracting the radii. 

We  studied the effects of  including the pionic interactions  
in previous work \cite{Cramer:2004ih,Miller:2005ji}.
 Distorted waves, instead of
 plane waves,  were used to represent  the pion wave functions.
The resulting formalism
is called the Distorted Wave Emission Function (DWEF) formalism 
because the emission function 
used to describe the space-time extent of the emitting system is 
dressed by the final state interactions. 
We found that  
it is possible to simultaneously describe the measured 
HBT radii and pionic spectra by including  
 the effects of pion-medium final state interactions obtained by solving the relevant
relativistic wave equation. 
These interactions are
so strongly attractive that the pions act as essentially massless objects
 inside the 
medium. The medium acts as if it is  free of the chiral condensate that 
is the source of the pion mass, and therefore 
 acts as a system with a restored chiral symmetry.  Other solutions of the HBT puzzle
have been proposed. See the review \cite{Lisa:2005dd}.

Some theorists have questioned 
whether  waves, produced by incoherent sources, unaffected by final state
interactions, interfere with those that are affected by 
final state interactions. That this interference occurs was demonstrated at least  
as early as 1979  \cite{Gyulassy}, again in the nineties \cite{Barz:1996gr,Barz:1998ce,hh,th},
and there have been two recent
publications confirming that conclusion \cite{Kapusta:2005pt}, \cite{Wong:2003dh}.
The former  derive a general formula for the correlation 
function of two identical particles  including 
 multiple elastic scatterings in the medium in which the two particles are produced.
Numerical results for the case of soft final state interactions are presented. 
Ref.~\cite{Wong:2003dh}
includes the effects occurring when  
emitted particles undergo multiple scattering with medium particles.
Using the Glauber theory of multiple scattering at high energies 
and the optical model at intermediate energies, it is found 
that multiple scattering leads to an absorption. 

Despite this progress, several conceptual issues remain. These include
understanding: the meaning of the imaginary part of the optical potential,
the role of the energy dependence of the optical potential \cite{Pratt:2007pf},
and the relationship between the sources producing the pions and the
optical potential. Therefore, we find it worthwhile to 
re-investigate the effects of quantum mechanical treatments of 
final state interactions. 

Our procedure is to
repeat the  derivation of Ref.~\cite{BP-HBT} using a simple Lagrangian. First,
the original plane wave treatment  is reproduced using our notation, Sect.~II. 
Then the
effects of a real optical potential are incorporated, Sect.~III.
  The result is a re-derivation of of the DWEF formalism. 
However, a deeper understanding is needed to correctly account for the effects of a
complex optical potential. This can only be incorporated using a coupled channels
formalism, Sect.~IV. We find that including the complete effects of an imaginary optical 
potential requires a modification to the DWEF 
formalism that is presently incalculable. However, the optical potential
used in \cite{Cramer:2004ih,Miller:2005ji} was dominated by its real part. 
In particular, in Sect.~V we find that setting the imaginary part optical potential to zero 
does not significantly change our description of the data. 
Sect.~VI is reserved for a summary and discussion.

\section{The Pratt Formalism}

The space-time extent of a source of pions can be inferred by measuring
 the pionic correlations known as  the Hanbury Brown-Twiss effect \cite{HBT,HBT1}. The correlation function function $C(\bfp,\bfq)$ is defined to be
\bea C(\bfp,\bfq)\equiv {P(\bfp,\bfq)\over P(\bfp)P(\bfq)},\eea where $P(\bfp_1,\cdots\bfp_n)$ is the probability of observing pions of momentum $\bfp_i$ all in the same event. The identical nature of all pions of the same charge cause $C(\bfp,\bfp)=2$. The width of the correlation function is related to the space-time extent of the source.

A state created by a random pion source $|\eta\rangle$
 is described by \cite{Gyulassy}
\bea |\eta\rangle=\exp[\int d^4x \eta(x)\gamma(t)\hat{\psi}^\dagger(x)]|0\rangle=\exp[\int d^3p dt \eta(\bfp,t)\gamma(t)
c^\dagger(\bfp)e^{-iE_pt}]|0\rangle, \label{etadef}
\eea where $\hat{\psi}^\dagger$ is the pion  creation operator in the Heisenberg representation,  $\gamma(t)$ is the random phase factor that 
takes the chaotic nature of the source into account, and $c^\dagger(\bfp)$ is the creation operator for a pion of momentum $\bfp$.. In particular, 
 an average over collision  events gives
\bea \langle \gamma^*(t)\gamma(t')\rangle=\delta(t-t'),\; \langle \gamma^*(t_1)\gamma^*(t_2)\gamma(t_3)\gamma(t_4)\rangle=\delta(t_1-t_3)\delta(t_2-t_4)+\delta(t_1-t_4)\delta(t_2-t_3).\label{random}\eea
We note that as written, the state $|\eta\rangle $ is not normalized to one. However, the normalization constant will divide out of the numerator and denominator of the correlation function.
Therefore we do not make the normalization factor explicit here, but 
note that it enters when we calculate the
pion spectrum.

For $\psiop$ and its time derivative to obey the Heisenberg commutation relation one has 
\bea\sqrt{E_pE_{p'}}[c(\bfp),c^\dagger(\bfp')]=\delta^{(3)}(\bfp-\bfp').\eea
Furthermore, we define
\bea \eta(\bfp,t)\equiv\int d^3x e^{-i\bfp\cdot\bfx}\eta(x).\label{ft}\eea
The state $|\eta\rangle$  is an eigenstate of the destruction operator in the Schroedinger representation, $c(\bfp)$:
\bea c(\bfp)|\eta\rangle=\int dt e^{iE_p t}{\eta(\bfp,t)\over E_p}\gamma(t)|\eta\rangle.\label{cex}\eea 
The correlation function is 
\bea C(\bfp,\bfq)={\langle\eta|c^\dagger(\bfp)c^\dagger(\bfq)c(\bfq)c(\bfp)|\eta\rangle\over
\langle\eta|c^\dagger(\bfp)c(\bfp)|\eta\rangle \langle \eta|c^\dagger(\bfq)c(\bfq)|\eta\rangle}
.\label{corr}\eea
The use of \eq{random} and \eq{cex} in the numerator of \eq{corr} yields
\bea \langle\eta|c^\dagger(\bfp)c^\dagger(\bfq)c(\bfq)c(\bfp)|\eta\rangle
=\langle\eta|c^\dagger(\bfp)c(\bfp)|\eta\rangle
\langle\eta|c^\dagger(\bfq)c(\bfq)|\eta\rangle +|\langle\eta|c^\dagger(\bfp)c(\bfq)|\eta\rangle|^2.\label{top}\eea
Furthermore
\bea \langle\eta|c^\dagger(\bfp)c(\bfq)|\eta\rangle =\int dt \exp[-i(E_p-E_q)t]{\eta^*(\bfp,t)\eta(\bfq,t)\over E_pE_q}.
\label{matel}\eea
The quantity  $g(x,\bfp)$ is denoted the emission function and is defined as 
\bea g(x,\bfp)=\int d^3x' \eta^*(\bfx+{1\over2}\bfx',t)\eta(\bfx-{1\over2}\bfx',t)e^{i\bfp\cdot\bfx'},
\eea
so that
\bea \int {d^3p\over(2\pi)^3} g(x,\bfp)e^{-i\bfp\cdot\bfz}=\eta^*(\bfx+{1\over2}\bfz,t)\eta(\bfx-{1\over2}\bfz,t)\\
 \int {d^3p\over(2\pi)^3} g((\bfy+\bfy')/2,t,\bfp)e^{-i\bfp\cdot(\bfy-\bfy')}=\eta^*(\bfy,t)\eta(\bfy',t).
\label{good}\eea
The second expression appears in the right-hand-side of \eq{matel}  
(if one uses \eq{ft}) so that we may write 
\bea\langle\eta|c^\dagger(\bfp)c(\bfq)|\eta\rangle =
\int d^4x {\exp[-i(p-q)\cdot x]\over E_pE_q}g(x,{(p+q)\over2}).\label{etag}\eea
Using \eq{etag} with $\bfp=\bfq$ shows that the function $g(x,\bfp)/E_p^2$ is the probability of emitting a pion of momentum $\bfp$ from a space-time point $x$. Using \eq{top} and \eq{etag} in \eq{corr} gives the desired expression:
\bea  C(\bfp,\bfq)=1+{\int d^4x\;d^4x'g(x,{1\over2}\bfK)g(x',{1\over2}\bfK)\exp[-ik\cdot(x-x')]\over\int d^4x\;d^4x'g(x,\bfp)g(x',\bfq)},\label{corr2}\eea where $\bfK\equiv\bfp+\bfq$ and $k\equiv(E_p-E_q,\bfp-\bfq)$, and the factors of ${1\over E_pE_q}$ have canceled out.

From a formal point of  view, a 
key step in the algebra is the relation between the Heisenberg representation pion creation operator
$\hat{\psi}^\dagger(x)$ and its momentum-space Schroedinger representation counterpart $c^\dagger(\bfp)$ that appears in \eq{etadef}:
\bea\hat{\psi}^\dagger(x)=\int d^3p \;c^\dagger(\bfp)\;{e^{-i\bfp\cdot\bfx}\over(2\pi)^{3/2}}\;e^{i E_p t}\label{complete}\\
\hat{\psi}(x)=\int d^3p \;c(\bfp)\;{e^{i\bfp\cdot\bfx}\over(2\pi)^{3/2}}\;e^{-i E_p t}
\eea
The operators $c^\dagger(\bfp)$ ($c(\bfp)$) are coefficients 
of a plane wave expansion for $\hat{\psi}^\dagger(x)\;(\psiop(x))$, with the
  plane wave functions  ${e^{i\bfp\cdot\bfx}/(2\pi)^{3/2}}$ being the complete set of basis functions.
However, one could re-write $\hat{\psi}^\dagger(x)\; (\psiop(x))$ as an expansion using   any set  of complete   wave functions. We shall exploit this feature below.

\section{Distorted waves --  Real Potential }

We represent the random classical source emitting pions that interact with a real, time-independent external potential ${\cal U}$ by the Lagrangian density:
\bea-{\cal L}=\hat{\psi}^\dagger (-\partial^2+{\cal U}+m^2)\psiop+j(x)\psiop.\label{lang}\eea 
The current operator $j(x)$ is closely related to the emission function $g$, 
\cite{Gyulassy}. In this Lagrangian
 the terms ${\cal U}$ and $j(x)$ are independent. Thus the relation between
the emission function   and ${\cal U}$  derived in \cite{pd} need not be satisfied.

The field operator $\psidagop$  can be expanded in the mode functions 
$\psi^{(-)}_\bfp$ that satisfy:
\bea  (-\nabla^2+{\cal U})\psi^{(-)}_\bfp(\bfx)=p^2\psi^{(-)}_\bfp(\bfx).\label{mode}\eea
These wave functions obey the usual completeness and orthogonality relations
\bea \int d^3p \psi^{(-)*}_\bfp(\bfx)\psi^{(-)}_\bfp(\bfy)=\delta^{(3)}(\bfx-\bfy)\label{comp}
\\
 \int d^3x 
\psi^{(-)*}_\bfp(\bfx)    
\psi^{(-)}_{\bfp^\prime}(\bfx)
=\delta^{(3)}(\bfp-\bfp'),\label{orth}\eea
so that one may use the field expansion
\bea \psiop(x)=\int d^3p\;\psi^{(-)}_\bfp(\bfx,t)e^{-iE_pt}d(\bfp),\label{dexp}\eea
with $d^\dagger(\bfp)$ being the creation 
operator for pions of momentum $\bfp$ in the basis of \eq{mode}.  
The expansion  \eq{dexp} assumes that ${\cal U}$ produces no bound states. 
If so, one the integral term would be augmented
by a term involving a sum over discrete states.

The availability of mode expansions when distortion effects are included
means that the simplification of the correlation function can 
proceed as in the previous section. We again use \eq{etadef} and \eq{random}.
The use of the field expansion \eq{dexp} enables a generalization of the function  $\eta(x)$:
\bea \eta(x)=\int d^3p \psi^{(-)*}_\bfp(\bfx,t)\tilde{\eta}(\bfp,t),\eea with
\bea\tilde{\eta}(\bfp,t)\equiv \int d^3x \psi^{(-)}_\bfp(\bfx,t)\eta(x),\eea
so that
\bea |\eta\rangle=\exp[\int d^3p dt \tilde{\eta}(\bfp,t)\gamma(t)
d^\dagger(\bfp)]|0\rangle. \label{etatilde}\eea The ability to obtain a
 relation between the $\tilde{\eta}(\bfp,t)$ and $\eta(x)$ rests on the relations \eq{comp} and \eq{orth}.

The state $|\eta\rangle$ is an eigenstate of $d(\bfp)$.  Thus the result
\bea C(\bfp,\bfq)=1+
{{|}\langle\eta|d^\dagger(\bfp)d(\bfq)|\eta\rangle{|}^2\over
\langle\eta|d^\dagger(\bfp)d(\bfp)|\eta\rangle \langle \eta|d^\dagger(\bfq)d(\bfq)|\eta\rangle},\label{corrnew}\eea
very similar to \eq{corr},
is  obtained.
We need the matrix elements appearing in the numerator and find
\bea\langle\eta|d^\dagger(\bfp)d(\bfq)|\eta\rangle =  
\int d^4x d^3x'  
{\exp[-it(E_p-E_q)]\over E_pE_q}\psi_p^{(-)}(\bfx)\psi_q^{(-)*}(\bfx')\eta(x)\eta(\bfx',t).\label{etag1}\eea
and the use of \eq{good} allows us to obtain
\bea\langle\eta|d^\dagger(\bfp)d(\bfq)|\eta\rangle = {1\over E_pE_q}\int dtd^3xd^3x' {d^3p'\over (2\pi)^3}e^{it(E_q-E_p)}
e^{-i\bfp'\cdot\bfx'}\psi_\bfp^{(-)}(\bfx+\bfx'/2)\psi_\bfq^{(-)*}(\bfx-\bfx'/2)g(x,\bfp').\nonumber\\
\label{newer}\eea
This result, which can be applied for $\bfp\ne\bfq $ and for $\bfp=\bfq$, specifies the evaluation of the correlation function
of \eq{corrnew} with the result 
\bea C(\bfp,\bfq)=1+{S(K,k)\over S(p)S(q)} \label{newcc}
\eea 
where \bea S(K,k)\equiv %
\int d^4xd^3x'{d^3p'\over(2\pi)^3} 
e^{it(E_q-E_{p'})} 
e^{-i\bfp'\cdot\bfx'} 
\psi_\bfp^{(-)}(\bfx+\bfx'/2)
\psi_\bfq^{(-)*}(\bfx-\bfx'/2)g(x,\bfp'),\label{corr3}\eea
and
\bea S(p)\equiv
\int d^4xd^3x'{d^3p'\over(2\pi)^3}e^{-i\bfp'\cdot\bfx'}
\psi_\bfp^{(-)}(\bfx+\bfx'/2)\psi_\bfp^{(-)*}(\bfx-\bfx'/2)g(x,\bfp').\eea
This expression is also the one that appears in  the DWEF formalism
 \cite{Cramer:2004ih,Miller:2005ji}.
One could use either \eq{corr2} or \eq{newcc} 
to analyze data, but the extracted 
space time properties of the source $\eta(x)$ would be different.

We need to 
comment on the possible momentum and energy dependence of the optical potential.
The completeness and orthogonality relations are obtained with any Hermitian 
${\cal U}$ which can therefore be momentum dependent, but not energy dependent.
As explained in Sect.~5 (Eq.~(43)) of Ref.~\cite{Miller:2005ji}, the real part of the 
potential  can and should be 
thought of as a momentum-dependent, but energy-independent potential.  If there were
true energy dependence a factor depending on the derivative of the potential with
respect to energy, \cite{Pratt:2007pf},  
would enter into the orthogonality and completeness relations.

\section{  Coupled channels}

The optical potential used in previous work  \cite{Cramer:2004ih,Miller:2005ji}
is complex. Using the necessary completeness and orthogonality relations to relate 
$\eta(x)$ to $\tilde{\eta}(\bfp,t)$  requires the use of a real potential. 
Therefore one needs to investigate possible corrections.

The optical potential or pion self-energy
is an effective interaction between the
pion and the medium. 
The medium is not an eigenstate of the Hamiltonian, but rather of $H_0$, which is
  the full
Hamiltonian minus the Hermitian operator representing the 
pionic final state interactions. 
 Eliminating the infinite number of possible states of $H_0$
and representing  these by a single state leads to a self-energy 
that is necessarily complex. Our procedure here is
to specifically consider the infinite number of states of the medium, 
obtain a Lagrangian density that involves  Hermitian interactions, and 
derive the optical potential formalism and any  corrections to it. 

Let $P_n$ denote a projection operator for the medium to be in a given eigenstate 
of $H_0$,  $n$. 
These obey 
\bea \sum_n P_n=1, P_nP_m=\delta_{n,m}P_n.\eea
For the case of
 $\pi$-nuclear scattering, $n$ would represent
the nuclear eigenstates. Here $n$ represents states of the medium in the absence of its interactions with pions. 
The correlation function is now given by 
\bea C(\bfp,\bfq)={\sum_n P_n(\bfp,\bfq)\over \sum_n P_n(\bfp)\sum_m P_m(\bfq)}\eea
where $P_n(\bfp)$ is the probability for emission of a pion of momentum $\bfp$ 
from the medium in a  
state $n$. Similarly $P_n(\bfp,\bfq)$ is the probability 
for emission of a pair of pions of 
momentum $\bfp,\bfq$ from the medium in a state $n$.  The sums over $n$ account for the inclusive nature of the process of interest.

It is convenient to 
 define the product of the field operator with the projection operator $P_n$:
\bea\psiop_n(x)\equiv \psiop(x) \;P_n,\eea with
\bea \psiop(x)=\sum_n \psiop_n(x),\label{newpsi}\eea using 
the complete nature of the set $n$. The Lagrangian density is given by
\bea-{\cal L}=\sum_{n}\partial\hat{\psi}_n^\dagger\cdot\partial\hat{\psi}_n
+\sum_{n,m}{\psi}_n^\dagger \left((m_\pi^2 +M^2_m)\delta_{nm} +{\cal U}_{nm}\right)\psiop_m+\sum_n j_n(x)\psiop_n(x)\label{lang1},\eea
where  \bea {\cal U}_{nm}={\cal U}^*_{mn}\equiv (\hat{\calU})_{nm}\eea 
and  $\hat{\calU}$ is the 
Hermitian interaction operator and $M_m^2$, the $m$ matrix element of the diagonal operator $M^2$, represents the effects of the different energies of the states labeled by $m$.
The field operator $\psiop_n$  can be expanded in the mode functions 
$\psi^{(-)}_{\bfp,n}$:
\bea  \sum_{m\ne n}{\cal U}_{nm}(\bfx)\psi^{(-)}_{\bfp,m}(\bfx,t)=(p^2+\nabla^2-M_n^2-\calU_{nn}(\bfx))
\psi^{(-)}_{\bfp,n}(\bfx,t).\label{mod1}\eea 
Here the potential ${\cal U}$ is taken as a local operator in the position space of the outgoing pion.

To see the correspondence between the formulation of \eq{lang1} and \eq{mod1},
let $\psiop_1$ correspond to the field operator (and state) of the previous section and solve formally for
 $\psi^{(-)}_{\bfp,m}$ in terms of $\psi^{(-)}_{\bfp,1}$.
It is convenient to define the operator $\widetilde\calU$ with matrix elements given by
\bea\tilde{\calU}_{n,n'} \equiv (1-\delta_{n,1})(1-\delta_{n',1})\calU_{n,n'}\eea
Then 
\bea \psi^{(-)}_{\bfp,n\ne1}=\sum_{m\ne1}\left({1\over \nabla^2+p^2-M^2-\widetilde{\calU}-i\epsilon}\right)_{nm}\calU_{m1}\psi^{(-)}_{\bfp,1},
\label{onec}\eea where $(\nabla^2+p^2-M^2)_{nm}\propto\delta_{n,m},$ and $M^2$ is an operator giving 
$M_n^2$ when acting on the state $n$.
 Then rewrite \eq{mod1} 
in  terms of $\psi^{(-)}_{\bfp,1}$ as 
\bea \calU_{11}\psi^{(-)}_{\bfp,1}+\sum_{m,n\ne1}\calU_{1n}\left({1\over \nabla^2+p^2-M^2-\calU-i\epsilon}\right)_{nm}
\calU_{m1}\psi^{(-)}_{\bfp,1}=(p^2+\nabla^2-M_1^2)\psi^{(-)}_{\bfp,1}\eea

The complex object $$\calU_{11}+\sum_{m,n\ne1}\calU_{1m}\left({1\over \nabla^2+p^2-M^2-\widetilde{\calU}-i\epsilon}\right)_{m,n}
\calU_{n1},$$ a non-local operator in coordinate space, 
 can be identified with the optical potential, given by the operator 
$V(Z)$ as a function of a complex variable $Z$:
\bea V(Z)=\calU_{11}+\sum_{m,n\ne1}\calU_{1m}\left({1\over \nabla^2+Z-M^2-\calU }\right)_{m,n}
\calU_{n1}.\label{vopt}\eea

We proceed by employing 
\eq{newpsi} and \eq{lang1} to compute the correlation function. The solutions of \eq{mod1} form a complete orthogonal set:
\bea \sum_n\int d^3p \psi^{(-)*}_{\bfp,n}(\bfx)\psi^{(-)}_{\bfp,n}(\bfy)=\delta^{(3)}(\bfx-\bfy)\label{comp1}
\\
 \sum_n\int d^3x \;
\psi^{(-)*}_{\bfp,n}(\bfx)
\psi^{(-)}_{\bfp',n}(\bfx)
=\delta^{(3)}(\bfp-\bfp').\label{orth1}\eea
The field expansion is now 
\bea\psiop(x)=\int d^3p \;\sum_n \; a(\bfp) P_n\psi^{(-)}_{\bfp,n}(\bfx)e^{-iE_pt},\eea
so that
\bea |\eta\rangle=\exp[\sum_n\int d^4x \eta_n(x)\gamma(t)\int d^3p\;a^\dagger(\bfp)P_n\psi^{(-)*}_{\bfp,n}(\bfx)e^{iE_pt}]\sum_m
|0,m\rangle\label{etatilde1},\eea 
where the state $|0,m\rangle$ is the pionic vacuum if the medium is in the state $m$, and $\eta_n(x)$ represents the  source for the state $n$. 
These state vectors obey the relations
\bea \langle 0,n|0,m\rangle =\delta_{n,m}=
\langle 0,n|P_n|0,m\rangle .\eea
Define  
\bea \eta_n(\bfp,t)
\equiv\int d^3x\;\eta_n(\bfx,t)\psi^{(-)*}_{\bfp,n}(\bfx),\eea 
so that
\bea |\eta\rangle=
\exp[\int d^3p dt \gamma(t)
\sum_n\eta_n(\bfp,t)a^\dagger(\bfp)P_ne^{iE_pt}]\sum_m|0,m\rangle,\eea 
\bea a(\bfp)|\eta\rangle=\int dt\;\gamma(t)\sum_n{\eta_n(\bfp,t)\over E_p}P_ne^{iE_pt}|\eta\rangle.\eea
The emission probability is given by
\bea E_pE_q\sum_n\langle\eta|a^\dagger(\bfp)P_n a(\bfq)|\eta\rangle=\sum_{n}\int dtd^3xd^3y\eta_n^*(\bfx,t)\eta_n(\bfy,t)
\psi_{\bfp,n}^{(-)*}(\bfx)\psi_{\bfq,n}^{(-)}(\bfy)\;e^{i(E_p-E_q)t}\eea or using \eq{good}
\bea E_pE_q\langle\eta|a^\dagger(\bfp)a(\bfq)|\eta\rangle=\sum_{n}\int dtd^3xd^3y\int d^3p'\;g_n((\bfx+\bfy)/2,t,\bfp')
e^{-i\bfp'\cdot(\bfx-\bfy)}
\psi_{\bfp,n}^{(-)*}(\bfx)\psi_{\bfq,n}^{(-)}(\bfy)\;e^{i(E_p-E_q)t},\nonumber\\\label{newnn}\eea 
where \bea   g_n(x,\bfp)=\int d^3x' \eta_n^*(\bfx+{1\over2}\bfx',t)\eta_n(\bfx-{1\over2}\bfx',t)e^{i\bfp\cdot\bfx'}.\eea
If pionic final state interactions are ignored, the term $\sum_n g_n$ enters and
this may be identified with the emission function, $g$ of previous sections.

The expression \eq{newnn} is the same as \eq{newer} except that now we sum over the channels $n$. These sums
 may be expressed in
terms of the optical model wave functions of \eq{onec}.
The term of \eq{newnn} with $n=1$ corresponds to the DWEF formalism, and the  terms
with $n>1$ are corrections. We provide an example of a correction term.
Suppose part of the imaginary part of the optical potential arises from a pion-nucleon interaction that makes an intermediate $\Delta$. Then a term corresponding to 
one of $n>1$ involves the emission of a pion from a nucleon that makes an 
intermediate $\Delta$.

It is difficult to assess the importance of the second term in a general way. 
The only obvious limit is that if states with $n>1$ are not excited then  
$Im({ V})$ of \eq{vopt} must vanish. Conversely, if 
 $Im({V})$=0, the states $n>1$ must be 
above the threshold energy and the propagators that appear in the  correction terms
correspond to virtual propagation over a  small distance with limited effect.

\section{Numerical Assessment of the Effects of $Im(V)$ in Refs.~\cite{Cramer:2004ih,Miller:2005ji}}
\begin{figure}
\includegraphics[width=18 cm]{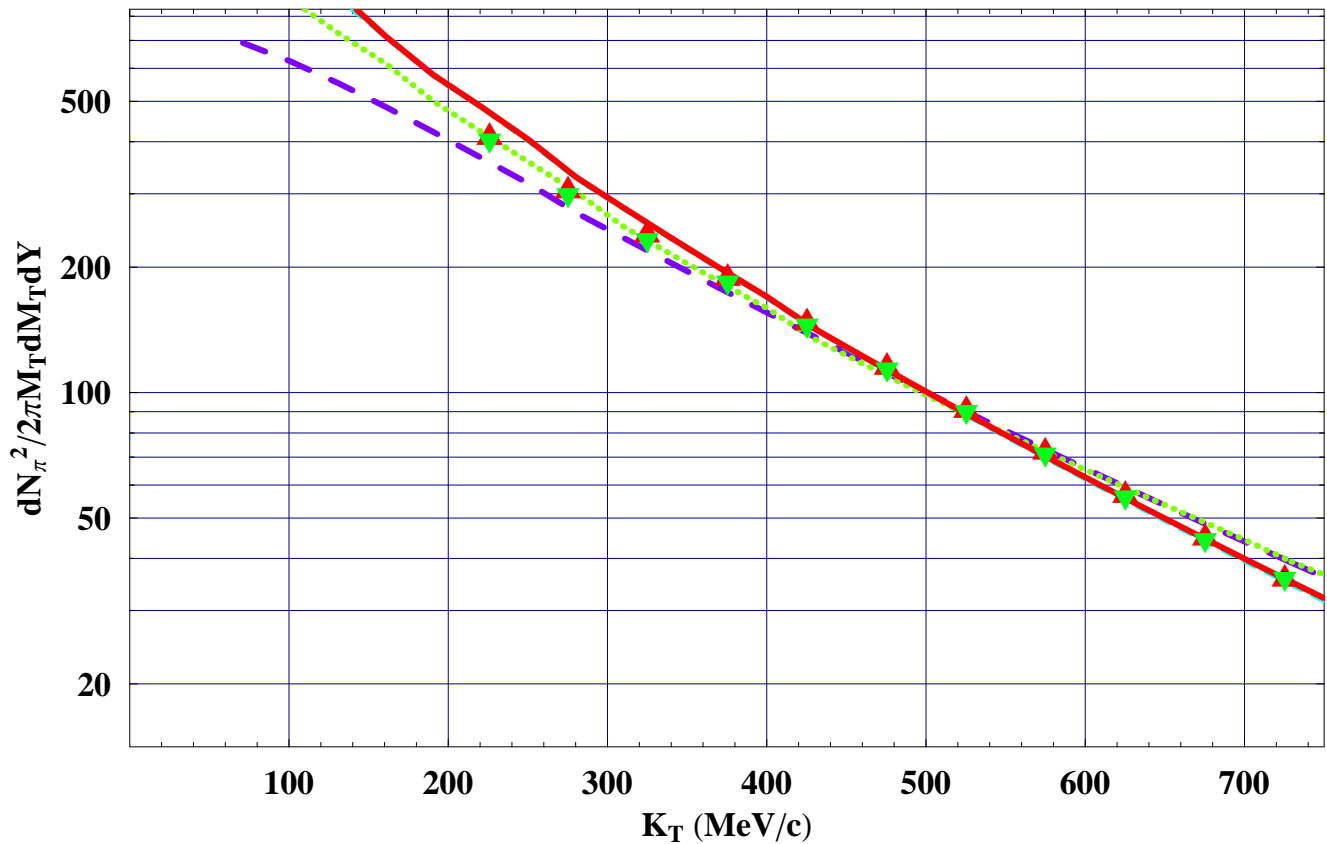}
\caption{\label{spec} Color online. Computed pionic spectrum.
Red upright triangles:  $\pi^-$ spectrum  (STAR)
Green inverted triangles:  $\pi^+$ spectrum points (STAR) \cite{Adams:2003xp}
Red solid line:  DWEF fit with vanishing 
imaginary part of the optical potential,
first line of Table I. 
Green dotted line: DWEF fit including search on the imaginary part of the optical
potential, 
second  line of Table I. 
Blue dashed line(almost entirely covered by the red solid curve): 
DWEF fit with vanishing 
imaginary part of the optical potential,$\Delta\tau$ =  1.5 fm/c, 
 third line of Table I.   
Violet long dashed line:  DWEF fit including search on $ \mu_\pi$, setting the optical
potential to essentially 0, 
fourth line of  Table I.
 }
\end{figure}

\begin{figure}
\includegraphics[width=20 cm]{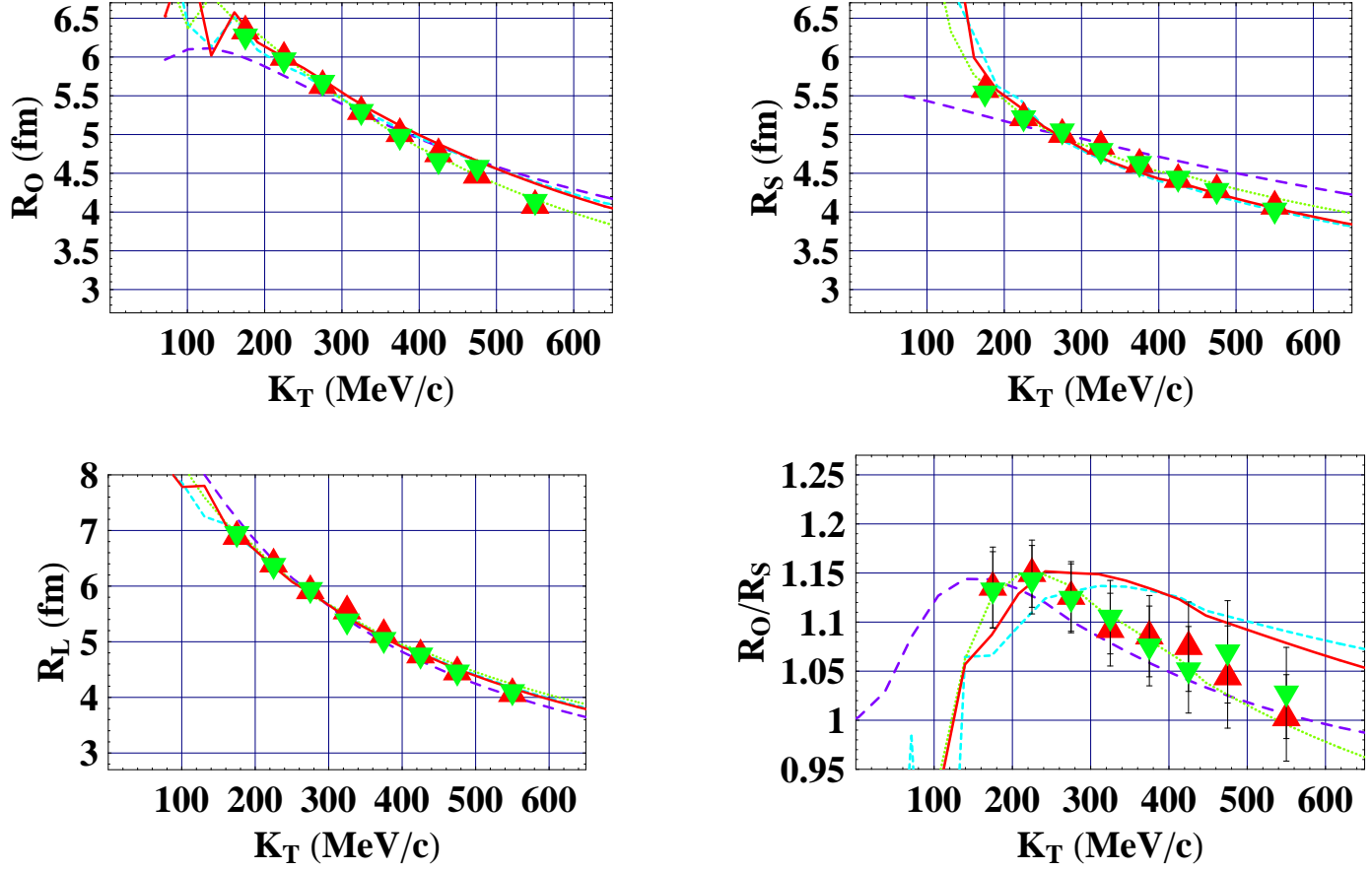}
\caption{\label{radii} Color online.  HBT radii. 
Curves are labeled as in Fig.~\ref{spec}. STAR data \cite{Adams:2004yc}}
\end{figure}

We proceed by assessing 
 the possible importance of the correction term for the work of
\cite{Cramer:2004ih,Miller:2005ji} by seeing what happens if the optical potential is 
taken to be purely real with no imaginary optical potential.  A variety DWEF fits are performed, see Table I. 
In \cite{Miller:2005ji}
the imaginary part of the 
optical potential as represented by the term $w_2$  is about 
one tenth of the real potential. It is therefore possible that, 
in the limit that $Im(w_2)=0$, there would be
no significant correction term, so we try to understand if removing the imaginary part of
the optical potential can be done without degrading the quality of the fit.
The results are shown in Figs.~\ref{spec} and \ref{radii}. 
An example of the previous calculations 
\cite{Cramer:2004ih,Miller:2005ji} is shown as the green dotted curve (second line of 
Table I). The red solid
curve (first line of Table I)
shows the result of setting the imaginary potential to a vanishingly small
value. 
This results in only 
a slightly worse description of the data.
The changes in the imaginary part of the optical potential 
$w_2$ are largely compensated by a reduction of the temperature 
from about 160 MeV to about 120 MeV. We also point out that 
 the length of the flux tube as represented by 
$\Delta \eta$ is vastly increased, providing greater justification to our previous
procedure of taking the length of the flux tube 
to be infinitely long in the longitudinal 
direction. However, the emission duration is reduced to
0 fm/c, which is similar to the results of the blast wave model
 \cite{Retiere:2003kf}. This means that all of the pionic emission occurs at a
single proper time. This value
justifies the use of a time-independent optical potential, but does 
seem to be difficult to understand because some spread of emission times
 is expected for a long-lived plasma. The results shown by the blue dashed curves 
(third line of Table I)
are obtained with fixing the emission duration to 1.5 fm/c, which is our previous
value \cite{Cramer:2004ih,Miller:2005ji}. The description of the  spectrum is basically
unchanged but 
 the
radii are less precisely described.
 The violet long-dashed curves (fourth line of Table I)
show the DWEF fit using a vanishing optical potential.
 This does not give a good description of the momentum dependence of the radii and 
is associate with the largest deviation between our calculations and the data as
represented by the $\chi^2$ values of Table~I.

It is clear that the precision of our description of the data is improved by
including the imaginary part of the optical potential. However, 
this is a quantitatively
but  not a qualitatively important effect. It is also true that including the real
part of the optical potential is a qualitatively important effect. These
 results suggest that the correction terms embodied by the terms with $n\ne1$ of 
\eq{newnn} are not very important, but non-negligible. It is also possible
that an optical potential with a different geometry than the volume
form that we have assumed might be able to account for the the neglected terms.
However, an accurate assessment  would require the  development 
a theory that involves dealing  with explicit  models for $g_n,j_n$ and $\calU$.

\begin{table*}
  \caption{Four  parameters sets 
obtained with slightly different procedures. These parameters are defined in \cite{Miller:2005ji}. The values of $\chi^2$ represents the accuracy of the description of the data.}
  \vspace{0.1cm}
  \begin{tabular}{|llcllllllll|}\hline
   $T(MeV)$ & $\eta_f$ & $\Delta\tau(fm/c)$ & $R_{WS}(fm)$ & $a_{WS}(fm)$& $w_0(fm^{-2})$& $~w_2$
& $\tau_{0}(fm/c)$ &  $\Delta\eta$ & $\mu_\pi(MeV)$&$\chi^2$\\
\hline
121   & 1.05  & 0 & 11.7 & 1.11 & 0.495 & 0.762 +0.0001$i$ & 9.20 &  70.7& 139.57&300\\
162 &   1.22 & 1.55& 11.9 &1.13 &0.488 &1.19+0.13$i$& 9.10&1.68&139.57&117\\
121   &  1.04 & 1.5 & 11.7 & 0.905 & 0.564 & 0.595 +0.0001$i$ &8.85  &70.7 & 139.57&451\\
144 &	0.990 &	2.07 &	12.57 &	0.876&	0.0001&	0.0001+	0.0001$i$	&6.85 &$\infty$& 83.5&1068\\
  \hline
      \end{tabular}
      \end{table*}
\section{Summary and Discussion}

 It seems clear from previous work including \cite{Gyulassy}-\cite{th} and 
the present Sect.~III that 
final state interaction effects on HBT interferometry are appropriately
 included by solving quantum mechanical
 wave equations. However, if the optical potential has an imaginary part, there
is an additional effect, embodied in  \eq{newnn} that needs to be included when
computing the emission probabilities and correlation function.   Thus the effects of
strong
quantum opacity must be accompanied by additional pion emission from the states
eliminated in the construction of the complex optical potential. In the
work of \cite{Cramer:2004ih,Miller:2005ji} the real part of the  optical
potential is very important and the imaginary part of the optical potential
is a small  effect. However, obtaining a similarly accurate reproduction
of the pionic spectra and HBT radii without this imaginary part causes the
emission temperature to drop from about 160 MeV to 120 MeV and the 
fitted emission duration time to drop to 0.
This indicates that,
in our model, either 
the final state interactions occur in the later times of the 
collision and that the emission occurs at only one proper time, or the inclusion of
emission from the states eliminated  in the construction of the 
complex optical potential is necessary.

\acknowledgments

This work was supported in part by the US Department of Energy 
under Grant No.~DE--FG02--97ER41014.

\end{document}